\RequirePackage{fix-cm}
\documentclass[12pt]{article}
\usepackage{color}
\usepackage{times}
\usepackage{color}
\usepackage{graphicx}
\usepackage{calc}
\usepackage{epsfig}
\usepackage{float}
\usepackage{amsfonts}
\usepackage[section]{placeins}
\usepackage{amsmath}
\usepackage{amssymb}
\usepackage[makeroom]{cancel}
\usepackage{bm}
\usepackage[utf8]{inputenc}
\usepackage{fancyhdr}
\usepackage{bbm}
\usepackage{xfrac}
\usepackage{lineno}
\usepackage{breakcites}
\usepackage{authblk}

\definecolor{purpura}{rgb}{0.5, 0.0, 0.5}
\definecolor{azul}{rgb}{0, 0.0, 0.6}
\definecolor{rojo}{rgb}{0.6, 0, 0}
\definecolor{verde}{rgb}{0, 0.6, 0}
\definecolor{turquesa}{rgb}{0, 0.5, 0.5}
\definecolor{marron}{rgb}{0.6, 0.4, 0}
\definecolor{gris}{rgb}{0.4, 0.4, 0.4}
\definecolor{celeste}{rgb}{0.5, 0.5, 0.8}
\definecolor{naranja}{rgb}{0.8, 0.5, 0}

\begin{document}

\begin{center}
{\large
{\bf Time- and frequency-resolved covariance analysis for detection and characterization of seizures from intracraneal EEG recordings}}

\vspace*{1cm}

\normalsize 
\textit{Melisa Maidana Capit\'an$^{12}$, Nuria C\'ampora$^{23}$, Claudio Sebasti\'an Sigvard$^1$, \\ Silvia Kochen$^{23}$ and In\'es Samengo$^{12}$}

\vspace*{1cm} 

{\footnotesize
1: Instituto Balseiro and Departamento de F\'{\i}sica M\'edica, Centro At\'omico Bariloche, Argentina. \\
2: Consejo Nacional de Investigaciones Científicas y Técnicas, Argentina. \\
3: Neurosciences and Complex Systems Unit (ENyS), Hospital El Cruce “Néstor Kirchner”, Universidad Nacional Arturo Jauretche, Argentina.}

\end{center}

\normalsize

\vspace*{0.7cm} 

\parskip 18pt
\baselineskip 0.25in

\begin{abstract}
The amount of power in different frequency bands of the electroencephalogram (EEG) carries information about the behavioral state of a subject. Hence, neurologists treating epileptic patients monitor the temporal evolution of the different bands.  We propose a covariance-based method to detect and characterize epileptic seizures operating on the band-filtered EEG signal. The algorithm is unsupervised, and performs a principal component analysis of intra-cranial EEG recordings, detecting transient fluctuations of the power in each frequency band. Its simplicity makes it suitable for online implementation. Good sampling of the non-ictal periods is required, while no demands are imposed on the amount of data during ictal activity. We tested the method with 32 seizures registered in 5 patients. The area below the resulting receiver-operating characteristic curves was 87\% for the detection of seizures and 91\% for the detection of recruited electrodes. To identify the behaviorally relevant correlates of the physiological signal, we identified transient changes in the variance of each band that were correlated with the degree of loss of consciousness, the latter assessed by the so-called Consciousness Seizure Scale, summarizing the performance of the subject in a number of behavioral tests requested during seizures. We concluded that those crisis with maximal impairment of consciousness tended to exhibit an increase of variance approximately 40 seconds after seizure onset, with predominant power in the theta and alpha bands, and reduced delta and beta activity.
\end{abstract}


{\bf Accepted in Biological Cybernetics, June 2020.}

\newpage

\section{Introduction}
\label{sec:intro}

The last decades have witnessed an explosion in computational techniques aimed at automatically detecting and characterizing epileptic seizures \cite{tzallas2012,orosco2013,alotaiby2014,ulatecampos2016,boubchir2017}. Some are based on simple measures, that assess the amount of power in different frequency bands \cite{bartolomei2008}. These methods are easily implemented, but their performance is limited, since a single dimension, as the ratio of high-to-low frequency power, is often not enough to encompass different types of seizures. Others focus on detecting some of the specific features that neurologists are trained to look for, when scrutinizing a recording \cite{harner2009,liu2013}. These methods can achieve a good detection performance, but are limited by the fuzziness of the definition of the targeted features, since there is no precise consensus among experts of what exactly constitutes a feature \cite{wilson2002}. Other approaches decide in terms of the visual properties of the EEG signal when plotted in the time-frequency domain \cite{boubchir2014,boubchir2015}, or of the total amount of synchrony \cite{schevon2007,warren2010,evangelista2015,courtens2016,bonini2016}, or of the coherence \cite{wang2017,aggarval2017}  or of the phase-to-amplitude coupling \cite{edakawa2016,liu2017,campora2019,dellavale2020}. Yet others \cite{kharbouch2011,liu2012,donos2015,heller2018,hugle2018}, make no assumptions on the distinctive features tagging seizures, and by means of supervised-learning algorithms, employ machine-learning techniques to discover the function that maps EEG signals to an output that distinguishes between “seizure” and “no seizure”. 

Many of these methods, though sometimes effective in their detection precision and recall, fit literally thousands of parameters, and are hard to interpret, since their outcome is not based on quantities on which clinical intuition is based. Neurologists analyze EEG data in terms of frequency bands, and they are able to associate specific features of the EEG signal with specific changes in the behavioral state of the patient. Therefore, when developing a tool for automatic analysis of EEG signals, it is important to first decide whether the aim is to produce an accurate algorithm for seizure detection, or rather, to design a tool that assists neurologists to identify the features that mark the onset of the seizure, to characterize its propagation, and to reveal the features that co-vary with the clinical manifestations. This paper ascribes to this second goal. We filter the EEG signal in the same frequency bands neurologists usually analyze them, and we propose an unsupervised seizure-detection algorithm based on covariance analysis. The approach is simple enough to be implemented online, and discloses the temporal properties of the EEG signal that are correlated with the degree of loss of consciousness, the latter, assessed by a standard behavioral test. 

Covariance analysis has long been used in neuroscience when analyzing electrophysiological recordings of the activity of single sensory neurons that are selective to specific stimulus features \cite{bryant1976,deruyter1988,simoncelli2004,samengo2013}. The physiological signal is typically discrete ({\em spike} or {\em no spike}) and the stimulus varies from trial to trial, testing the effect of a collection of features that are are the candidate relevant dimensions to which the neuron might be sensitive to. The stimulus dimension along which spiking probability is modulated maximally is identified as the most relevant one. We here extend this technique to the study of time-continuous EEG signals, and search for the temporal structures that are maximally modulated by seizures. We later use the same idea to identify the temporal features in the signal that are modulated by the degree of loss of consciousness in each seizure.


\section{Materials and Methods}
\label{sec:mm}


\subsection{EEG data sets}
\label{sec:mm1}

Long-term intra-cranial EEG recordings were obtained from 5 hospitalized patients during 24-hour video-EEG monitoring, lasting for 5 days. The electrodes were implanted as a pre-surgery evaluation, and their anatomical targeting was decided for each patient on the base of available non-invasive information about the localization of the epileptogenic zone. The ictal clinical semiology was obtained from the videos of seizures. EEG signals were obtained at 2 kHz sampling rate from 32 seizures of variable duration (mean: 78 seconds, SD: 41 seconds) recorded from 5 patients with hypothesis of temporal epilepsy, each patient with a different number of implanted macro electrodes (Ad-Tech depth electrodes, 0.86mm diameter, 5mm contact spacing, between 5-10 contacts).  The total number of analyzed segments was 1599. Each segment contributing to this total number corresponds to a single patient, a single crisis, and a single recording site on a single electrode. Segments also included non-epileptic activity, both before and after the seizure. On average, the seizure took 9\% (SD 5\%) of each segment. The study has been performed in accordance with the ethical standards as laid down in the 1964 Declaration of Helsinki and its later amendments, and was approved by the ethics committee of the El Cruce Néstor Kirchner and the Ramos Mejía Hospitals. All patients signed a written informed consent form before their voluntary participation in the study.


\subsection{Identification of epileptic seizures by expert evaluation}
\label{sec:mm2}

Two experts trained and experienced in video-EEG interpretation reviewed all video-EEG recordings. Each seizure was reviewed 3 to 4 times in its entirety to identify pathological signs. The onset of a seizure was established at the first electroencephalographic change. The termination of the seizure was ascribed to the moment when rhythmic activity concluded, the EEG showed a diffused attenuation or slowing, or more than 90\% of the EEG channels were slow and the patient's stereotyped behavior ceased.


\subsection{Assessment of the degree of loss of consciousness}
\label{sec:mm3}

In 29 of the 32 crisis, the degree of loss of consciousness was assessed by a clinical evaluation performed by a trained examiner \cite{campora2016}. The result was summarized by a value on the Consciousness Seizure Scale (CSS) \cite{arthuis2009}, derived on the base of the degree and adequacy of responsiveness of the patient, visual awareness, identification of the seizure as such, and the degree and type of induced amnesia. The index varies between 0 and 9, with 0-1 representing full awareness, 2-5 moderate consciousness impairment, and 6-9, severe loss of consciousness. 


\subsection{Representation of the signals}
\label{sec:rep}

\begin{figure}[ht]
\centerline{\includegraphics[keepaspectratio=true, clip = true,
scale = 0.7, angle = 0]{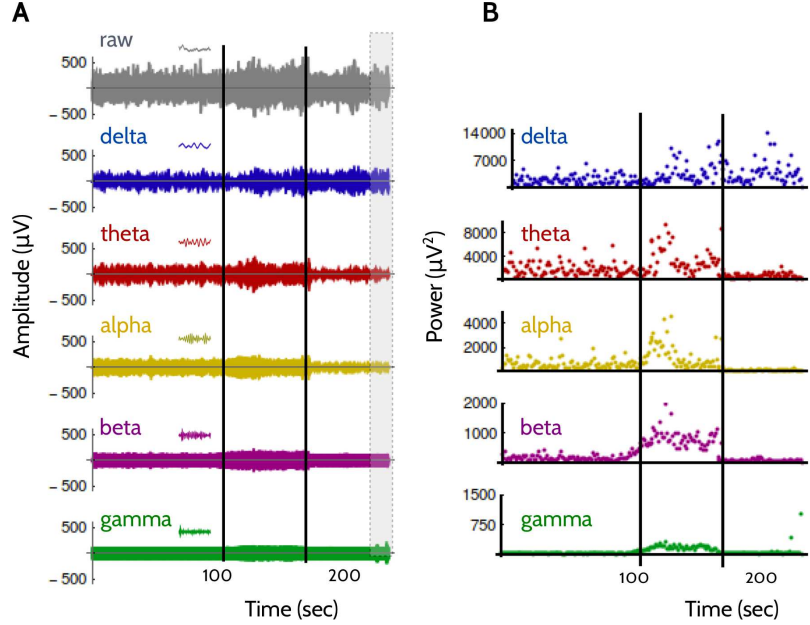}} \caption{\label{f:1}
{\bf Representation of the EEG signal}. A: Voltage traces. Top: raw recorded signal obtained in a given contact. Rows 2-6: Same signal filtered in different frequency bands. Vertical lines: Initiation and termination of the ictal activity. Insets: Detail of the signal in a 2-second window outside the ictal period (seconds 88-90). Gray band on the left: temporal span of the 19-second window used to sample each covariance matrix. B: Power of each filtered signal as a function of the position of the 1-second sliding window. The seizure is more evident on the right panel, which represents the variance of the signals on the left.}
\end{figure}
Neurologists typically inspect EEG signals visually, often filtering specific frequency bands. Figure~\ref{f:1}A shows the raw signal (top) obtained from a given recording site on a given macro-electrode, and the same signal filtered in different frequency bands. We represented the signal as a 5-dimensional vector 
\[
\mathbf{s}^t = (s_\delta, s_\theta, s_\alpha, s_\beta, s_\gamma),
\]
where the supra-index $t$ stands for vector transposition, and the components  representing the voltage traces in the delta (1-3.4 Hz), theta (3.4-7.4 Hz), alpha (7.4-12.4 Hz), beta (12.4 - 24 Hz) and gamma (24 - 97 Hz) bands, respectively. The variance of each filtered signal in a one-second non-overlapping sliding window is shown in Fig.~\ref{f:1}B. 


\subsection{Covariance analysis}
\label{sec:cov}

We define the mean vector $\langle \mathbf{s} \rangle$ as the temporal average of the whole collection of vectors $\mathbf{s}(t)$, and $C_0$ as the $5 \times 5$ covariance matrix with entries $(C_0)_{ij} = \langle s_i \ s_j\rangle - \langle s_i \rangle \ \langle s_j \rangle$. The matrix $C_0$ can be taken to its diagonal form $D$ by a coordinate transformation $D = O^t \ C_0 \ O$, where $O$ is an orthogonal matrix. Following \cite{samengo2013}, in order to spot departures from the normal state, all sampled vectors $\mathbf{s}$ are transformed into vectors $\mathbf{s}' = D^{-\sfrac{1}{2}} O \mathbf{s}$. The new coordinates are here referred to as the {\it symmetric} ones. The total covariance matrix of the entire collection of points once transformed to the symmetric coordinates is the unit matrix.

To detect the seizures, we defined a sliding window lasting for 19 seconds, that was shifted along the entire recording, one second at a time. For each position $t$ of the window, the 19 symmetric data points contained in it were used to calculate a local covariance matrix $C(t)$. The largest eigenvalue of this local matrix defined a temporal sequence $\lambda(t)$ capturing the variance of the data along the direction of maximal variance inside each 19-second temporal window. Significantly increases of $\lambda(t)$ above unity are candidates to tag the onset of seizures. In order to avoid false positives due to transient fluctuations induced by limited samples, the departure was required to last for at least an interval $\tau$. This requirement was instantiated by defining a temporally filtered signal 
\[
\bar{\lambda}(t) = \int_{t - \sfrac{\tau}{2}}^{t + \sfrac{\tau}{2}} \lambda(t') \ {\rm d}t'.
\]
Seizures were detected at those points in time where $\bar{\lambda}(t)$ surpassed a given threshold.The value of $\tau = 54$ s was chosen to maximize the agreement between our criterion and that obtained from the trained neurologists.


\section{Results}
\label{sec:results}


\subsection{An unsupervised method based on PCA of the power in each band}
\label{sec:res1}

Our first goal is to develop an unsupervised method for detecting seizures based on features that physicians are accustomed to work with. First, the signal is filtered in the delta, theta, alpha, beta and gamma frequency bands (Sect.~\ref{sec:rep}). Covariance analysis is based on a principal-component analysis of the filtered signals. Our first aim is to identify the direction in 5-dimensional space in which the variance displays an anomalous behavior when entering into the seizure. Our second aim is to reveal the dimension in which the variance varies systematically with the degree of loss of consciousness.

\begin{figure}[ht]
\centerline{\includegraphics[keepaspectratio=true, clip = true,
scale = 0.7, angle = 0]{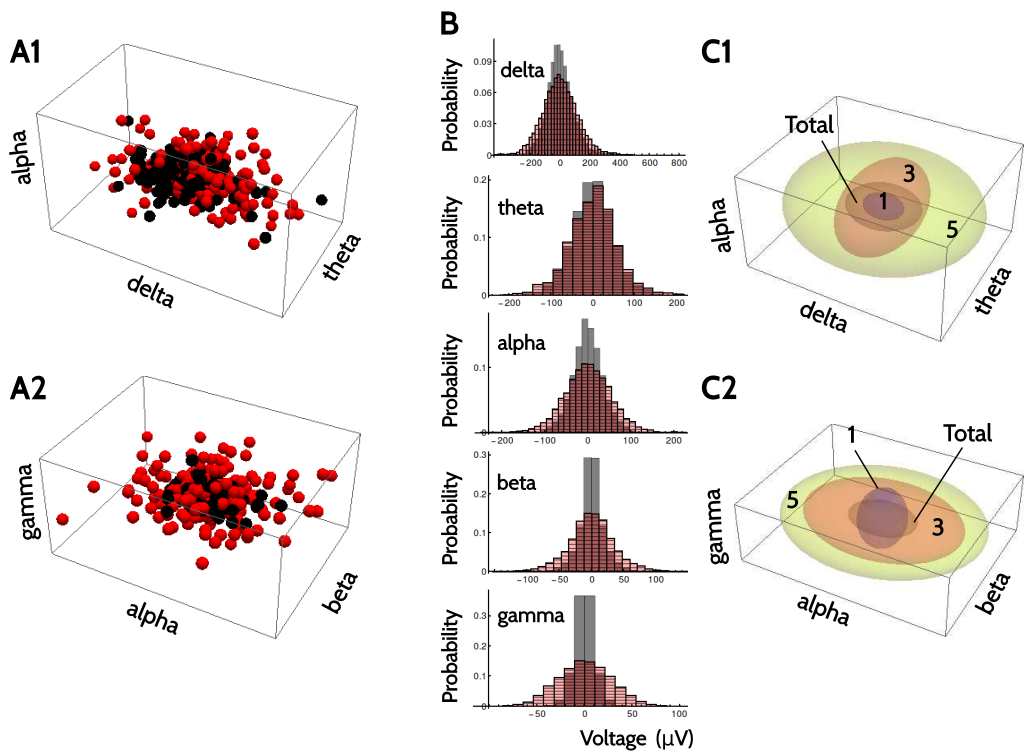}} \caption{\label{f:2}
{\bf 5-dimensional representation of the EEG signal}. A: Distribution of filtered signals of Fig.~\ref{f:1}A. A1: Voltages of the delta, theta and alpha components. A2: Variances of the alpha, beta and gamma components. Black: Non-ictal period. Gray (red online): Ictal period. The same number of ictal and non-ictal data points is shown, though the ictal voltages have a smaller variance, so the tend to appear more tightly clustered around the center of the distribution. B1: Voltage histograms of each component in ictal (striped, pink online) and non-ictal (gray) temporal windows. C: Ellipsoids describing the contour lines of the best Gaussian fit to the data of 3 selected seizures of a single patient, in the delta, theta, alpha (C1) and the alpha, beta, gamma (C2) subspaces. The ellipsoid marked as ``Total'' fits the whole collection of data, including ictal and non-ictal activity. Numbers indicate the CSS of each crisis.}
\end{figure}

In Fig.~\ref{f:2}A we show the distribution of voltages of the $5$-dimensional vectors $\bm{s}^t$ inside and outside the seizure. The low- and high-frequency (A2) components are displayed in different panels (A1 and A2, respectively) for better visualization. The $\alpha$ component appears in both.

In Fig.~\ref{f:2} A, the gray points (red online) correspond to the ictal period, and their values occupy a larger region of space than the black points (non-ictal), most particularly in panel A2. The variance of the ictal distribution, hence, is larger than the non-ictal one, and the difference is most evident in the high-frequency subspace (lower panel), where the black points cluster at the center, and the red ones, lie at the periphery. Our goal is to design an algorithm that can identify ictal from non-ictal activity by detecting transient increases in the variance. To that end, in Fig.~\ref{f:2}B we depict the ictal and non-ictal voltage distributions. The two conditions are most clearly differentiated in the higher frequency bands, from alpha to gamma (Fig.~\ref{f:2}B, lower panels). From this example, one would be tempted to conclude that an adequate method to detect ictal activity should be based on identifying anomalous increases in the variance in the high-frequency bands. This conclusion, however, does not take into account the considerable patient-to-patient variability in the statistical properties of seizures, nor the variability from crisis to crisis in one given patient. Although the epileptogenic zone \cite{bancaud1965} (the neural network responsible for seizure onset) is always the same for a given subject, the propagation dynamics may vary from one seizure to the next. Therefore, there need not be such thing as “a typical seizure”. Fig.~\ref{f:2}C shows the contour ellipsoids corresponding to the best Gaussian fit of the data points of 3 different seizures from the same patient. The gamma dimension is indeed convenient to identify one of the seizures (marked with “1” in Fig.~\ref{f:2}C), but this result cannot be generalized to other seizures. Based on this evidence, we here relinquish the aspiration to identify the seizures by their behavior in some fixed frequency band. Instead, we focus on characterizing the statistical properties of the non-epileptic state. The aim is to construct a reliable description of the so-called ``normal'' distribution, and then identify the seizures as dramatic departures from normality, irrespective of the direction in which the abnormality manifests itself. This strategy ensures that even a collection of seizures with highly variable properties can be identified, as long as they all clearly deviate from the non-epileptic state.

In the full recording, the non-epileptic state contains a number of samples that is overwhelmingly larger than that of the epileptic state. In our data, in which all segments contain a crisis, seizures comprise about 9\% of the signal. The statistical properties of the non-ictal activity, hence, can be determined even when analyzing the whole collection of data, containing both the ictal and the non-ictal time segments, since the latter vastly dominate. As a consequence, we may calculate the first two moments of the normal state using the entire recorded signal. 

\begin{figure}[ht]
\centerline{\includegraphics[keepaspectratio=true, clip = true,
scale = 0.7, angle = 0]{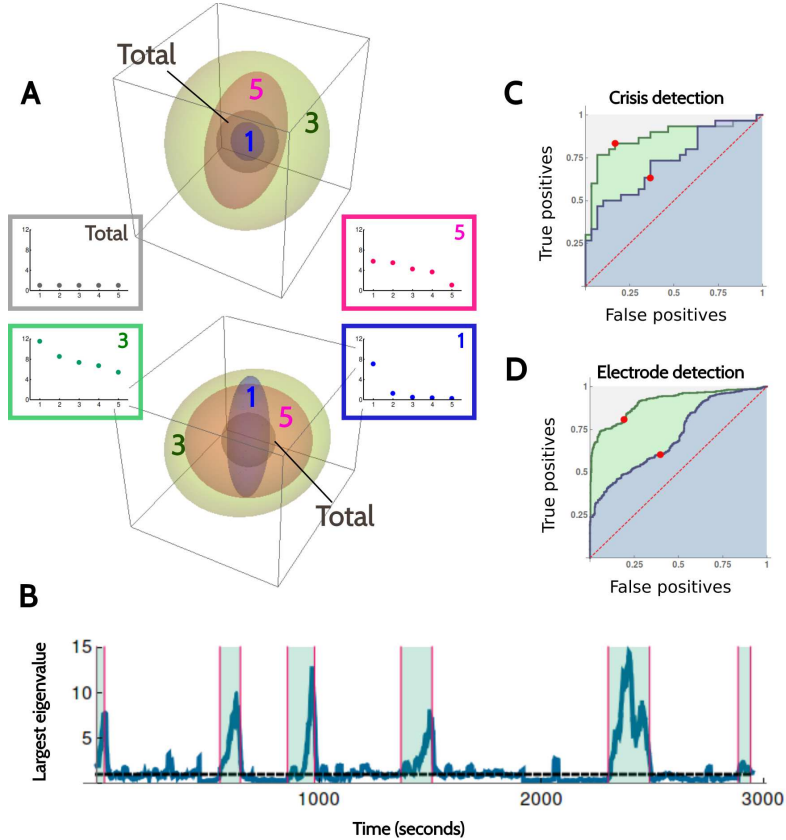}} \caption{\label{f:3}
{\bf Detection of seizures as departures from the normal state}. A: Ellipsoids describing the contour lines of the best Gaussian fit to the data of each crisis, in the symmetric space. In A1, the coordinate axes are the first three eigenvectors of $C_0$, and in A2, the $3^{\rm rd}, 4^{\rm th}$ and $5^{\rm th}$. The inner gray sphere fits the whole collection of data, including ictal and non-ictal activity, and the associated distribution has unit variance in all directions. Numbers indicate the CSS of each seizure. Insets depict the spectra of eigenvalues of a local matrix encompassing each crisis. Each eigenvalue is the square of the length of one of the principal axes of the corresponding ellipsoid. B: Largest eigenvalue of the local matrix, as a function of the position of the sliding window. Vertical lines indicate the initiation and termination of each crisis, as diagnosed by a trained neurologist. The dashed horizontal line marks the unit variance. C and D: Receiver-operating characteristic (ROC) curves with the performance of the detection algorithm in identifying seizures (C) and recruited electrodes (D), as compared with the ground truth provided by neurologists. The dashed line represents random detection. Green and blue areas: With and without transformation to the symmetric space.}
\end{figure}

The covariance matrix $C_0$ of the whole collection of vectors $\mathbf{s}(t)$ can be easily calculated (Sect.~\ref{sec:cov}). The eigenvectors of $C_0$ are the principal directions of the ellipsoid that encompasses the best Gaussian fit of the entire recording, and the eigenvalues are the variances of the data along these directions. We have verified that typically, the eigenvectors are aligned with the coordinate axes, and the eigenvalues decrease monotonously from the delta to the gamma dimensions. In other words, in normal circumstances, both the mean value and the variance along the low-frequency components are much larger than the mean and the variance along the high-frequency components. This property  is also evident in the range covered by the horizontal axes of Fig.~\ref{f:2} B. Following \cite{samengo2013}, in order to spot departures from the normal state, we make a coordinate transformation that turns the ellipsoidal distribution of the normal state into a spherical distribution, with unit variance in all directions. The resulting symmetric distribution is seen as a gray sphere in Fig.~\ref{f:3}A, and the regions of space occupied by each crisis, as elongated (colored online) ellipsoids.

In order to detect the time windows where the data departed significantly from the normal state (gray sphere of unit radio in Fig.~\ref{f:3}A), we defined a sliding window lasting for 19 seconds, that was shifted along the entire recording, one second at a time. For each position of the window, the filtered signals (once transformed to the symmetric space) were used to calculate a local covariance matrix. If the window fell on a non-ictal segment, the eigenvalues of the local matrix should be approximately unity, coinciding with the spherical distribution--with some unavoidable fluctuations, due to limited sampling. As the window enters into a seizure, the eigenvalues are expected to differ from unity, since as shown above, seizures correspond to collections of data points whose variance deviates from the normal state. In this paper, we propose to identify the onset of an epileptic seizure in a single channel as the point in time in which the largest eigenvalue of the local matrix is significantly larger than unity. In order to avoid false positives due to local fluctuations, the deviation of the anomalous eigenvalue was required to last for a time interval of duration $\tau$ (technical details in Sect.~\ref{sec:cov}). The seizure was assumed to last for as long as the local matrix sustained the significant departure. 

Figure~\ref{f:3}B shows the temporal evolution of the largest eigenvalue of the local matrices computed in each of the positions of the sliding window. The maximal eigenvalue increases dramatically during the epileptic seizures diagnosed by trained neurologists (shaded rectangles). The algorithm detects a seizure when the eigenvalue crosses a pre-defined threshold, and remains above it for an interval of duration $\tau = 54$ seconds. As the threshold is increased, the receiver-operating characteristic (ROC) curves of Figs.~\ref{f:3}C and D are traversed from right to left. The local nature of the detection procedure can be used to identify the onset of the crisis in each of the recording channels, and thereby provide a spatio-temporal description of the propagation of the seizure.


\subsection{Performance of the detection algorithm}
\label{sec:res2}

Assuming that the detection performed by trained neurologists is the ground truth, receiver-operating curves (ROC) can be constructed to assess the performance of the algorithm. ROC curves are constructed by plotting the number of true positives as a function of the number of false positives for each value of the detection threshold. A given crisis is detected when at least one of the recording channels indicated by physicians produces an eigenvalue that surpasses the chosen threshold inside the temporal window marked by physicians. A given electrode is identified if it produces an outlier eigenvalue inside the temporal window marked by neurologists. Figures~\ref{f:3}C and D display the ROC curves thus obtained.

The performance of the algorithm is assessed by the area beneath the ROC curve. The optimal operating point is the value of the threshold for which the precision (number of true detections / total number of detections) equates the recall (number of true detections / total number of crisis).  Table~\ref{t1} summarizes the effectiveness of the detection procedure. We have included the performance of the method when the traces are not symmetrized, to highlight the relevance of the normalization step.

Of all the algorithms discussed in the literature, the only one that has been previously formulated solely in terms of the amount of power in different frequency bands, that can be implemented online, and that has been reported in intra-cranial signals is the so-called epileptogenic index (EI) \cite{bartolomei2008}, in which seizures are detected when the power in the high-frequency bands increases with respect to that of low-frequency bands. We therefore compared the performance of our algorithm with that of the EI using our corpus of signals. The results are also included in Table~\ref{t1}, for comparison. The main differences between the two methods are that (a) our method makes no a priori assumptions about the frequency bands in which seizures manifest themselves, and (b) our method requires a transformation to the symmetric space, so that the asymmetry of the normal state is evened out.

Quite remarkably, the seizures that our algorithm fails to identify are those in which all eigenvalues are notably {\it smaller} than unity, implying that there are some seizures in which the power in all frequency bands is abnormally small. As far as we know, this is a novel result. Unfortunately, those seizures cannot be detected by identifying the eigenvalues that are smaller than a given threshold, because in the normal state, small eigenvalues accumulate near zero. Hence, detecting seizures by pinpointing abnormally small eigenvalues produces a huge number of false positives. 

\begin{table}\caption{\label{t1} Comparison between the performance of the method proposed here (Normalized covariance) with the same method without normalization (Na\"{\i}ve covariance), and with the Epileptogenic Index (EI)}.
\begin{center}
\begin{tabular}{ccccc}
Detection of & Method & Area below ROC curve & True Positives & False Positives \\ \hline
Crisis & Normalized covariance & $87$\% & $83$\% & $17$\% \\
Electrodes & Normalized covariance & $91$\% & $81$\% & $19$\% \\
Crisis & Na\"{\i}ve covariance & $73$\% & $63$\% & $37$\% \\
Electrodes & Na\"{\i}ve covariance & $71$\% & $60$\% & $40$\% \\
Crisis & EI & $75$\% & $70$\% & $30$\% \\
Electrodes & EI & $71$\% & $60$\% & $40$\%
\end{tabular}
\end{center}
\end{table}


\subsection{Transient characteristics observed during the loss of consciousness}
\label{sec:res3}

\begin{figure}[ht]
\centerline{\includegraphics[keepaspectratio=true, clip = true,
scale = 0.6, angle = 0]{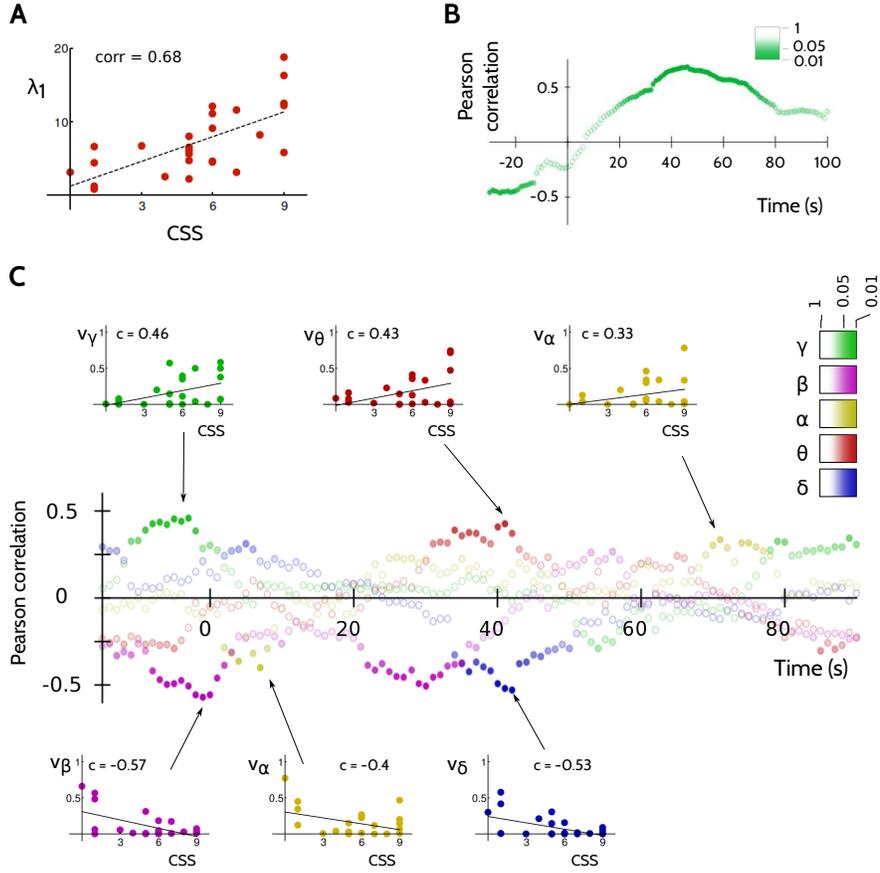}} \caption{\label{f:4}
{\bf Correlation between eigenvalues and CSS}. A: Average of the magnitude of the largest eigenvalue of all the local covariance matrices corresponding to all recording channels involved in each seizure as a function of the CSS. Matrices were located at the 46th second after seizure onset, as detected by our algorithm. The Pearson correlation coefficient is 0.68 (significantly different from zero, $p = 7 \times 10^{-5}$). B: Pearson correlation coefficient (calculated as in panel A) as a function of the temporal location of the window used to calculate the local covariance matrices, measured with respect to the onset of the crisis, as determined by our algorithm. The saturation of each data point represents the $p$-value obtained from a null hypothesis of uncorrelated variables (inset). C: Temporal evolution of the Pearson correlation coefficient between the CSS and the absolute value of each component of the eigenvector associated with the largest eigenvalue (after transforming back to the original space and normalizing), averaged between all the electrodes recruited by the seizure. The saturation of each data point represents the $p$-value obtained from a null hypothesis of uncorrelated variables (inset). Scatter plots: Data from which the chosen Pearson correlation values were calculated.}
\end{figure}

For each local matrix, the magnitude of the largest eigenvalue is a measure of the degree of departure from normality. We therefore verified whether such magnitude co-varied with the degree of loss of consciousness. In Fig.~\ref{f:4}A we see that both quantities are indeed correlated, and in Fig.~\ref{f:4}B, the correlation is shown to be maximal at approximately 50 seconds after seizure onset.

Interestingly, the seizures that the algorithm fails to detect have small CSS values. Since all the eigenvalues of these crisis are small, if these crisis were included in Fig.~\ref{f:4}A, the correlation between the magnitude of the largest eigenvalue and the CSS would be even larger.

The eigenvectors associated with the largest eigenvalue, when transformed to the original space, exhibited several transient features that were significantly correlated with the degree of loss of consciousness (Fig.~\ref{f:4}C). Inside the temporal window $(40-60)$ seconds after seizure onset, when also the eigenvalue was significantly correlated with the CSS (Fig.~\ref{f:4}B), the eigenvector associated with seizures with severe consciousness impairment tended to have particularly large components in the theta band, and to a lesser degree, also in the alpha band. In addition, the power in the delta and the beta bands was significantly reduced. 


\section{Discussion}
\label{sec:conc}

Many of the methods proposed so far for discriminating ictal from non-ictal activity require extensive training iterations, due to the fact that the distinctive features of epileptic seizures vary markedly from patient to patient, and sometimes, even from crisis to crisis in a single patient. Here we propose to identify the seizures only by diagnosing a significant deviation of the variance of the distribution of the filtered signals from the so-called ``normal'' distribution, no matter the direction in which the anomaly arises. The performance of the detection algorithm is around 90\%.  Although more precise methods exist \cite{tzallas2012}, our method has the advantage of being fully transparent, of requiring no training, of being amenable to online implementation, and of imposing no requirements on the amount of data during the epileptic state.

The crucial ingredient is the variance of the different frequency bands. In the normal state, the delta band is the one with largest variance. For the method to work, the transient amount of power in each band has to be compared with the baseline power. From the methodological point of view, this means to work in the so called {\em normalized} or {\em symmetric} space, where the variance of different frequency bands in the inter-ictal state is evened out. In the symmetric space, “normality” appears spherical, and seizures are detected as departures from normality. 

Deviations in the variance of slow and intermediate frequencies becomes relevant when assessing the degree of loss of consciousness, as revealed by the components of the most relevant eigenvector. Several studies have started to characterize the spatio-temporal features of those seizures that diminish or abolish consciousness \cite{bonini2016,campora2016,arthuis2009,blumenfeld2012}. In particular, in \cite{blumenfeld2012}, loss of consciousness was associated with increased synchronization in the alpha band. In addition, an increase of power in the theta band has been previously associated with states of minimal consciousness \cite{blumenfeld2012,schiff2014}. Hence, some previous studies seem to indicate that intermediate frequency bands are correlated with impaired consciousness. Our method confirms this result, since loss of consciousness is maximal in those seizures with large power in the range of 3-12 Hz, some 50 seconds after seizure onset. 

The components of the eigenvectors in different bands fluctuate rapidly, implying a complex dynamics, dominated by transient phenomena. The prominence of each band lasts for only a few tens of seconds, a time scale that is slower than the induced loss of consciousness. Seizures are therefore confirmed to be a highly non-stationary process, structured into several phases. 

Our analysis reveals that those seizures with significant compromise of consciousness are represented by eigenvectors that display larger amounts of theta power 30-50 seconds after seizure onset. To a lesser degree, the same effect was observed with the amount of alpha power. Contrastingly, in the same temporal window, the delta and the beta components of the eigenvector manifested the opposite trend: Higher CSS was associated with lower power. To relate these findings with the present theories of consciousness, we note that temporal lobe complex partial seizures often involve abnormal theta spike-wave activity \cite{blumenfeld2012}. The {\sl network inhibition hypothesis} postulates that loss of consciousness is induced by selective inhibiting subcortical arousal systems leading to depressed function of the higher order association cortex, including the default-mode network areas \cite{danielson2011}. As a consequence, slow activity emerges, with the typical pattern observed in slow-wave sleep \cite{englot2010}. The association between loss of consciousness and high theta + alpha components reported here, if confirmed in a larger population of patients, is compatible with the hypothesis that initial focal activity in the temporal lobe triggers anterior thalamic nuclei \cite{norden2002} to produce polyspike epileptic discharges that propagate the seizure to several cortical regions, and simultaneously inducing ventral posteromedial nuclei to produce sleep-like spindles interrupting thalamo-cortical information flow \cite{feng2017}. Our results, however, do not confirm the hypothesis that the amount of delta power can serve to discriminate between seizures with high and low compromise of consciousness, at least not with intracraneal electrodes, after averaging over all recruited channels. Visual inspection of seizures with different CSS values reveals that delta power evolves significantly throughout the crisis, but is not exclusively enhanced in those with high CSS. Therefore, though slow activity may indeed be required to inhibit the default-mode network, as hypothesized by other studies \cite{englot2010}, our results suggest that, if such is the case, some other source of slow activity is likely to exist, which is present in seizures with low CSS.

The alpha rhythm has long been known to be suppressed during the acquisition of bottom-up sensory information \cite{berger1929}, as well as during semantic information processing \cite{klimesch1997,klimesch1999}, directed attentional shifts \cite{sauseng2005,sauseng2011}, and conscious stimulus detection \cite{romei2007}. Indeed, during the normal, physiological state, alpha synchronization is often seen as a functional correlate of inhibition of cognitive and motor tasks \cite{klimesch2007}. In turn, beta activity has long been associated with sensory events and information processing during wakefulness. Our results, in which alpha power increases and beta power diminishes during the seizures that induce unconsciousness, is consistent with these findings. 

Finally, in the late 80-ies and early 90-ties, several studies proposed that consciousness requires the synchronization of populations of neurons via rhythmic discharges in the gamma range \cite{cauller1988,gray1989,crick1989}. Our results do not confirm a prominent role of gamma power, at least, not in the time window where the anomalous variance is maximal. This negative result is aligned with more recent findings, in which gamma power is more linked to the process of reporting conscious information processing, than to actually experiencing it \cite{koch2016,redinbaugh2020}. 

To our knowledge, this is the first study that not only reports the performance of an algorithm in the detection of the crisis, but also, in the detection of the individual electrodes involved in the onset and the propagation of the crisis, and to correlate the variance of each frequency band with the CSS. We therefore believe that the method has the potential to characterize the temporal, the spatial and some of the cognitive properties of seizures. 

\section*{Acknowledgements}

This work was supported by Agencia Nacional de Investigaciones Cient\'{\i}ficas y T\'ecnicas PICT Ra\'{\i}ces 2014 N. 1004, Consejo Nacional de Investigaciones Cient\'{\i}ficas y T\'ecnicas PIP 0256.

\section*{Conflict of interest}

The authors declare that they have no conflict of interest.




\end{document}